\documentstyle[prl,multicol,aps,epsf]{revtex}
\begin{document}
\draft

\title{Hole Localization in Underdoped 
Superconducting Cuprates Near 1/8th Doping} 
\author{J. L. Cohn and C. P. Popoviciu} 
\address{Department of Physics , University of Miami, Coral Gables, FL 33124}
\author{Q. M. Lin and C. W. Chu}
\address{Texas Center for Superconductivity at University of Houston and the
Department of Physics, Houston TX 77204}

\maketitle

\begin{abstract}
Measurements of thermal conductivity versus 
temperature over a broad range of
doping in YBa$_2$Cu$_3$O$_{6+x}$ and 
HgBa$_2$Ca$_{n-1}$Cu$_n$O$_{2n+2+\delta}$ ($n$=1,2,3)
suggest that small domains of localized holes 
develop for hole concentrations near $p$=1/8.  The data
imply a mechanism for localization that is intrinsic
to the CuO$_2$-planes and is enhanced via pinning associated 
with oxygen-vacancy clusters.

\end{abstract}

\begin{multicols}{2}
\narrowtext

There has been considerable recent interest in 
novel charge- and spin-ordered phases
which may compete with superconductivity in the 
cuprates\cite{Cuprates}.
This stripe order is favored at CuO$_2$ planar hole concentrations 
near $p$=1/8 for which the modulation wavelength is commensurate 
with the lattice, and when an appropriate pinning potential is present.  
Inelastic neutron scattering results\cite{DynamicStripes}
suggest that stripe modulations are
disordered and/or fluctuating in 
La$_{1-x}$Sr$_x$CuO$_4$ (La-214) and YBa$_2$Cu$_3$O$_{6+x}$ (Y-123).
Nuclear magnetic and quadrupole resonance (NMR and NQR) 
studies\cite{Hammel,HamScal} indicate the presence of localized holes 
in the CuO$_2$ planes.  Whether localized holes are a general feature of 
cuprates and contribute to short-ranged charge/spin segregation 
and the normal-state pseudogap\cite{Loram,PseudoGap} are 
fundamental issues of current interest. 

Here we report the observation of anomalies at $p$=1/8 
revealed in the oxygen doping dependence
of thermal conductivity ($\kappa$) in Y-123 and Hg cuprates.  Our principal
finding is that a fraction of the doped holes in the CuO$_2$ planes 
become localized near this particular doping level.  This fraction correlates with 
the oxygen vacancy concentration in the charge reservoir layers, consistent
with the formation of nanoscale, hole-localized domains that are pinned
near oxygen-vacancy clusters.

\begin{figure}
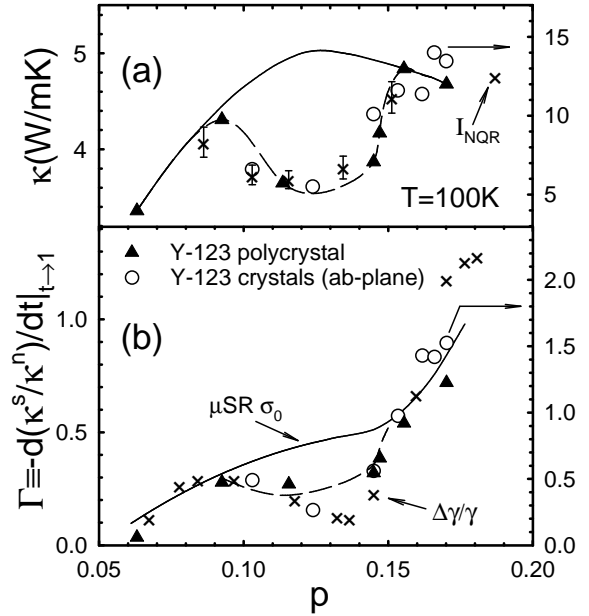

\caption{(a) Doping dependence of the normal-state $\kappa$ at $T$=100K 
for Y-123 polycrystals and the ab-plane of single crystals from 
Ref.~\protect{\onlinecite{Popoviciu}}. Values of $p$ were determined 
from thermopower measurements following Ref.~\protect{\onlinecite{PfromTEP}}
Also plotted ($\times$'s, right ordinate) is $(0.08+0.16I_{NQR})^{-1}$, 
where $I_{NQR}$ is the anomalous planar $^{63}$Cu NQR signal 
(Ref.~\protect{\onlinecite{HamScal}}, see text).  The dashed curve is a 
guide to the eye.  
The solid curve is discussed in the text. (b) The normalized slope 
change in $\kappa(T)$ at $T_c$ {\it vs} doping for each of the 
Y-123 specimens in (a).  
Also shown are the normalized electronic specific heat jump, 
$\Delta\gamma/\gamma$,
from Ref.~\protect{\onlinecite{Loram}} ($\times$'s), and the 
$\mu$SR depolarization rate (in $\mu$s$^{-1}$) 
from Ref.~\protect{\onlinecite{muSRY123}} (solid curve), 
divided by 1.4 and 2.7, respectively,
and referred to the right ordinate. The dashed curve is described in the text.}  
\label{KofPGammaofPY123}
\end{figure}

Two prominent characteristics of the in-plane heat conductivity in 
superconducting cuprates\cite{UherRev} are a normal-state $\kappa$ 
predominantly of lattice origin, and an abrupt increase in $\kappa$ 
at $T\leq T_c$.  To motivate our proposal that $\kappa$ 
probes localized holes, we demonstrate for Y-123 that 
its magnitude and temperature derivative at $T_c$ correlate 
with independent measures of local structural distortions (planar Cu NQR) 
and the superfluid fraction (specific heat jump), respectively, 
throughout the underdoped regime ($p\leq 0.16$). Consider data from 
previous work\cite{Popoviciu} that illustrate the $p$=1/8 features 
in $\kappa$ (Fig.~\ref{KofPGammaofPY123}).  
The upper panel shows normal-state data, $\kappa(p,T$=100K), for
an Y-123 polycrystal and the ab-plane of single crystals.  The 
lower panel shows, for the same specimens, 
the dimensionless slope change in $\kappa$ at $T_c$ defined as\cite{Popoviciu}
$\Gamma\equiv -d(\kappa^s/\kappa^n)/dt|_{t=1}$, where $t=T/T_c$ and 
$\kappa^s (\kappa^n)$ is the thermal conductivity in 
the superconducting (normal) state. $\Gamma$ measures the change 
in scattering of heat carriers (electrons and phonons) 
induced by superconductivity.  Both quantities exhibit local minima near 
$p$=1/8.

Figure~\ref{KofPGammaofPY123}~(a) implies a 
sharply reduced lattice conductivity ($\kappa_L$) about
$p$=1/8, since the electronic thermal
conduc-\break
\centerline{\epsfxsize=3.25in\epsfbox[50 60 540 770]{hgfig1.psc}}
\vskip -1.4in
\noindent
tivity ($\kappa_e$) at $T>T_c$ 
in optimally doped, single-crystalY-123
represents only $\sim$10\% of the total 
$\kappa=\kappa_e+\kappa_L$\cite{Krishana}. 
We attribute this suppression in $\kappa_L$ to phonon scattering by
local distortions of the CuO polyhedra.
In support of this interpretation is the doping behavior of the
relative intensity ($I_{NQR}$) of 
anomalous planar $^{63}$Cu NQR peaks for Y-123, indicative of 
charge inhomogeneity in the planes, and attributed to the 
presence of localized holes\cite{HamScal}.  
The simple assumption that the total thermal resistivity is a sum 
of a term proportional to this disorder ($\propto I_{NQR}$), 
and a constant term from other scattering, reproduces 
$\kappa(p)$ quite well [crosses, Fig.~\ref{KofPGammaofPY123}~(a)].
Studies of $\kappa$ in lightly-doped insulating cuprates and structurally 
similar manganites\cite{CohnManganites} provide further
evidence that bond disorder associated with localized charge 
is an effective phonon scattering mechanism in perovskites.

The electronic 
specific heat jump for Y-123\cite{Loram}, $\Delta\gamma/\gamma(p)$ 
[crosses, Fig.~\ref{KofPGammaofPY123}~(b)], 
correlates remarkably with $\Gamma(p)$.
Though the origin of the slope change in 
$\kappa$ at $T_c$ (i.e. electronic or phononic) 
has been a subject of some debate, this correlation
demonstrates unambiguously that $\Gamma$, like $\Delta\gamma/\gamma$, 
measures the difference between the normal-
and superconducting-state low-energy electronic spectral weight.  
This conclusion is 
robust since the correlation holds for both polycrystals
and single crystals; in spite of additional thermal resistance 
due to $c$-axis heat flow and granularity in the polycrystal, 
changes with doping reflect the intrinsic behavior of
the in-plane heat conduction.  

The muon spin rotation ($\mu$SR) depolarization rate 
($\sigma_0$)\cite{muSRY123}, 
a measure of the superfluid density, matches the data for $\Gamma$
and $\Delta\gamma/\gamma$ at $p<0.09$ and $p>0.15$ when suitably scaled
[solid curve in Fig.~\ref{KofPGammaofPY123}~(b)].   The sharp rise in 
$\sigma_0$ for $p>0.15$ is due to the superconducting condensate on 
oxygen-filled chains\cite{muSRY123}.  The range of $p$ over
which $\Gamma$ and $\Delta\gamma/\gamma$ deviate from the $\mu$SR curve 
{\it coincides} with the range of suppressed $\kappa$.  
The suppressed
transfer of spectral weight below $T_c$ is
consistent with the localization of a fraction of planar holes.
That $\sigma_0$ is not also suppressed near $p$=1/8
indicates that hole-localized domains do not 
inhibit the formation of a flux lattice in adjacent regions where holes 
are itinerant.  The scenario we have outlined resolves the 
prior inconsistency of the nonmonotonic  
$\Delta\gamma/\gamma$ doping behavior with the continuously rising pseudogap 
energy scale, inferred from numerous experiments 
in the underdoped regime\cite{Loram,PseudoGap}.

If our interpretation is correct the maximum density of localized holes
occurs at $p=1/8$.  This suggests a possible connection with the physics
of stripe formation and raises questions about the generality of this phenomenon
and the role of oxygen vacancies.  To address these 
issues we have studied $\kappa$
in HgBa$_2$Ca$_{m-1}$Cu$_m$O$_{2m+2+\delta}$ [Hg-1201 ($m$=1), Hg-1212 ($m$=2),
Hg-1223 ($m$=3)].  A single HgO$_{\delta}$ layer
per unit cell contributes charge to $m$ planes in 
Hg-12($m$-1)$m$ so that the oxygen vacancy concentration, $1-\delta$, 
increases with decreasing $m$ (at optimum doping, $\delta_{opt}\simeq0.18$ 
(Hg-1201), 0.35 (Hg-1212) and 0.41 (Hg-1223)\cite{Occupancy}). 

The preparation of the polycrystalline 
starting materials is
described elsewhere\cite{Samples}. Transport measurements were 
performed on specimens ($\sim1\times 1\times 3$mm$^3$) cut from as-prepared 
disks; each was measured repeatedly up to ten times 
after successive heat treatments at 
300-350$^{\circ}$C in flowing argon or vacuum to 
reduce the oxygen content. The thermopower ($S$) 
and thermal conductivity were measured 
during each experimental run
using 25$\mu$m type-E thermocouples and a steady-state
technique.  Heat losses via radiation and 
conduction through the leads (10-15\% near 300K 
and 1-3\% for $T<120$K) were determined in separate
experiments and the $\kappa$ data corrected. 
Uncertainty in the contact geometry introduces a $\pm 5$\% 
inaccuracy in $\kappa$.  

\begin{figure}
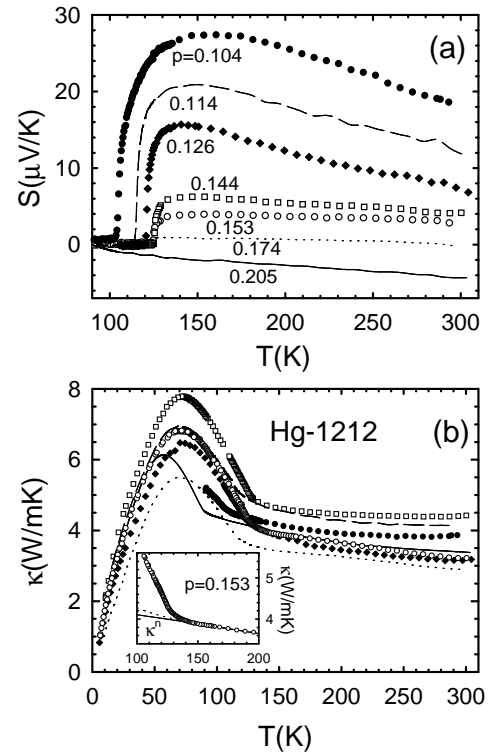

\caption{(a) Thermoelectric power and (b) thermal conductivity 
{\it vs} temperature for Hg-1212.  Data at 
$p=0.121$ and $p=0.132$ are omitted from both (a) and (b) for clarity.  
The inset shows $\kappa(T)$ near $T_c$
and extrapolated $\kappa^n$ curves for $p=0.153$.}
\label{TEPandKappa1212}
\end{figure}

\begin{figure}
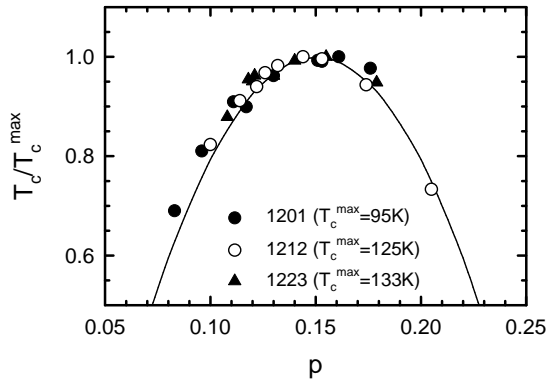

\vskip -1.58in
\caption{$T_c(p)$ normalized to $T_c^{max}$ for each of the Hg compounds.  
The solid
curve is $T_c/T_c^{max}=1-82.6(p-0.15)^2$.}
\label{TCofP}
\end{figure}

Figure~\ref{TEPandKappa1212} shows $S(T)$ and $\kappa(T)$ for the 
Hg-1212 specimen.  The thermopower curves are labeled by $p$, as 
determined from $S(290$K) and the universal $S(p)$ relations established
by Tallon {\it et al}\cite{PfromTEP}.  Data for Hg-1201 and Hg-1223
yield similar sets of curves.  Figure~\ref{TCofP} shows
$T_c(p)/T_c^{max}$ for each of the materials studied.  The data 
follow the generic high-$T_c$ phase\break 
\centerline{\epsfxsize=3.29in,\epsfbox[50 0 540 790]{hgfig2.psc}}
\vskip -1.15in
\break
\centerline{\epsfxsize=3in,\epsfbox[50 200 540 800]{hgfig3.psc}}
\noindent
curve\cite{PfromTEP},
approximated well by the parabolic form, $T_c(p)/T_c^{max}=
1-82.6(p-0.15)^2$ (solid curve, Fig.~\ref{TCofP}).

\begin{figure}
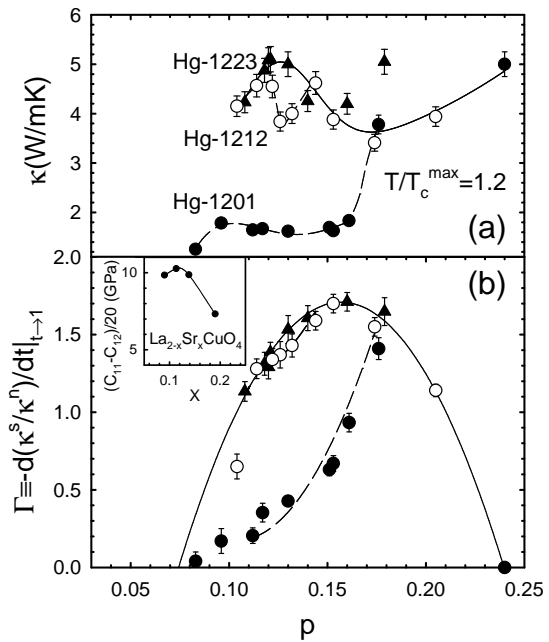

\vskip -.7in
\caption{(a) Doping dependence of the normal-state $\kappa$ at $T/T_c^{max}=1.2$ 
for each of the
Hg compounds.  Dashed curves are guides to the eye and the solid curve is
described in the text. 
(b) The normalized slope change in $\kappa(T)$ at $T_c$ 
{\it vs} doping for each of the Hg compounds at the 
same doping levels as in (a).  The solid line is $1.71-250(p-.157)^2$.  
Dashed curves are described in the text.  The inset shows elastic constant
data from Ref.~\protect{\onlinecite{LSCOelastic}}.}  
\label{KofPGammaofP}
\end{figure}

The doping dependencies, $\kappa(p,T/T_c^{max}$\ =\ 1.2) and $\Gamma(p)$ 
for each material are shown in Fig.~\ref{KofPGammaofP} (a) and (b), 
respectively.  As for the Y-123 data\cite{Popoviciu}, 
$\kappa^n$ was determined by polynomial fits to the
normal-state data at $t\geq 1.2$ (see inset, Fig.~\ref{TEPandKappa1212}). 
Uncertainties in the heat-loss corrections and extrapolations of 
$\kappa^n$ in the range $0.9\leq t<1.2$ are reflected in the error 
bars for $\Gamma$. Uncertainties in the extrapolations were established by varying 
the range of data and polynomial order used for fitting.

The data for the Hg materials are qualitatively similar to that of Y-123,
but with the magnitude of suppression near $p$=1/8 in $\kappa$ and $\Gamma$ 
increasing with the oxygen vacancy concentration.  Consider first
the $\Gamma(p)$ data for Hg-1223 which falls on an inverted parabola,
centered near $p=0.16$ [solid curve, Fig.~\ref{KofPGammaofP} (b)]. 
This behavior is similar to the $\mu$SR curve for Y-123, and 
suggests that the superconducting condensate is not suppressed in Hg-1223.
An extension of our interpretation for Y-123 implies that very few holes are 
localized in this compound.
The $\Gamma$ data for underdoped Hg-1201 and Hg-1212 are suppressed relative 
to that of Hg-1223 and maximally so near $p$=1/8, consistent with 
the localization of a fraction of holes.

A similar behavior is seen in the $\kappa$ data.   
For $p\leq 0.16$, Hg-1212 has $\kappa(p)$ indistinguishable from
that of Hg-1223, except in the narrow range of $p$ near 1/8 where it 
is sharply suppressed.  $\kappa$ for Hg-1201 is substantially 
suppressed over a broader doping range.
The implication is that in the absence of 
localized holes, $\kappa$ for underdoped Hg-1201 and Hg-1212 would follow 
the solid curve in Fig.~\ref{KofPGammaofP} (b), an interpolation through 
a composite of data for all three compounds, excluding the
data point at $p\simeq0.18$ for Hg-1223
and ranges of $p$ for Hg-1212 and Hg-1201 where $\kappa$ is suppressed. 
The increase in $\kappa(p)$ in the overdoped regime ($p\geq 0.16$) for all 
compounds is presumably 
due to the rising $\kappa_e$, and possibly to a rising $\kappa_L$ as well, 
the latter due, e.g., to a decrease in the phonon-electron scattering.  

Our data suggest that the maximum in $\kappa(p)$ 
observed near $p$=1/8 for Hg-1223
is an intrinsic feature of $\kappa_L$ in underdoped cuprates.
From kinetic theory, $\kappa_L=1/3C_Lv^2\tau$, where $C_L$ is the lattice 
specific heat, $v$ is the sound velocity, and $\tau$ the phonon relaxation time.
Interestingly the behavior of the normal-state elastic constants 
($\propto v^2$) of single-crystal La-214\cite{LSCOelastic} parallels that
of $\kappa$ for Hg-1223.  A substantial 
hardening of the lattice in the underdoped regime, with a maximum near 
$p=x\simeq$1/8, is observed for all symmetries (shear mode data are shown 
in the inset, Fig.~\ref{KofPGammaofP}).  An increasing $v^2$ is expected 
quite generally from elastic coupling to conduction electrons 
for a decreasing electronic density of states (the pseudogap)\cite{VsoundMax}.  
The appearance of a maximum suggests a competing effect, possibly related to 
the increasingly magnetic character of the planes.  Thus the maximum in 
$\kappa$ for underdoped Hg-1223 is plausibly attributed to a renormalized 
dispersion, and its suppression in compounds with localized holes to increased
scattering.  In spite of the oversimplification implicit in kinetic theory, 
these observations suggest that in the 
absence of localized holes, $\kappa(p)$ for Y-123 would have a similar behavior
[solid curve,  Fig.~\ref{KofPGammaofPY123} (a)].

Our interpretation dictates a proportionality between the
suppression of $\kappa$ and $\Gamma$, allowing for a self-consistency
check.  The lattice thermal re-\break
\centerline{\epsfxsize=3in,\epsfbox[30 20 600 800]{hgfig4.psc}}
\noindent
sistivity ($W\simeq 1/\kappa$) introduced 
by randomly-distributed, hole-localized domains should be 
proportional to their volume fraction since $\kappa$ is a bulk probe.
We then expect $\Gamma$ (and $\Delta\gamma/\gamma$) to be suppressed 
by an amount proportional to the same fraction, i.e. 
$\Delta\Gamma(p)=a\Delta W(p)$, where $\Delta\Gamma$ and $\Delta W$ are 
the amounts by which the measured $\Gamma$ and $W$ differ from values given by 
the solid curves in Fig.'s~\ref{KofPGammaofPY123} and~\ref{KofPGammaofP}. 
This simple relation agrees very well with the data [dashed curves in
Fig.'s~\ref{KofPGammaofPY123}~(b) and~\ref{KofPGammaofP}~(b)]
using $a^{1212}\simeq 1.7$W/mK, $a^{1201}\simeq 2.6$W/mK, and 
$a^{Y-123}\simeq 2.8$W/mK.  
  
The Hg data make it clear that oxygen vacancies play only a supporting role
in the localization of holes. 
The importance of 1/8th doping implies that the phenomenon
involves an excitation of the CuO$_2$ planes that is commensurate with 
the lattice.  A plausible candidate is a  
small domain of static stripe order\cite{Cuprates}, nucleated via pinning by a 
vacancy-induced mechanism.  Recent Raman scattering studies of 
optimally-doped Hg compounds\cite{HgRaman} implicate
vacancy clusters in pinning: the oscillator strength of the 
590 cm$^{-1}$ mode, 
attributed to c-axis vibrations of apical oxygen in an 
environment of four vacant nearest-neighbor dopant sites,
scales with  $\Delta\kappa$ and $\Delta\Gamma$.  Oxygen vacancies 
in both Y-123 and Hg cuprates\cite{Occupancy,Jorgensen} 
displace neighboring Ba$^{2+}$ toward the 
CuO$_2$ planes.  Resulting local distortions
of the CuO polyhedra could pin a charge stripe as do the octahedral
tilts in (La,Nd)-214\cite{Cuprates}.  Alternatively, a magnetic mechanism is 
possible.  In the Hg cuprates the shift of Ba atoms associated with a cluster 
of four vacancies nearest a CuO polyhedron will induce a positive potential 
in the planes that inhibits its occupation by a hole, thereby suppressing 
spin fluctuations and possibly fixing a Cu$^{2+}$ spin at the site.  
Larger vacancy clusters, surrounding adjacent CuO
polyhedra oriented along $\langle 1 0 0\rangle$ directions, may   
induce several spins to order antiferromagnetically via superexchange.
The presence of this spin-chain fragment would favor 
charge/spin segregation that is characteristic of stripe order. 

Consider the relevant length scales.
If randomly distributed, static stripe domains are to scatter 
phonons, their in-plane extent must be less 
than the phonon mean-free-path, $\Lambda$, and their mean separation 
comparable to $\Lambda$.
Expressing $\Lambda^{-1}$ as a sum
of terms for scattering by these domains and by all other processes, 
$\Lambda^{-1}=\Lambda^{-1}_{str}+\Lambda^{-1}_{other}$, we use the in-plane
thermal resistivity for Y-123 and kinetic theory to find\cite{NoteonMFP}
$\Lambda_{str}\simeq [3/C_Lv\Delta W(p$=$1/8)]\simeq 
70{\rm\AA}$ as an estimate of the separation between domains at $p$=1/8.
We can make a rough estimate of the fractional area of the planes 
having static stripes by taking the typical domain size to be 
$2a\times8a$ ($a$ is the lattice constant), the stripe unit cell  
suggested\cite{Cuprates} for (La,Nd)-214.  This yields a fraction
$16(a/\Lambda_{str})^2\simeq 0.05$.  The $\Gamma$ data suggest that 
this fraction is somewhat higher in Hg-1201, but apparently below the 
2-D percolation threshold of 0.50.  This presumably explains 
why no substantial $T_c$ suppression is observed near $p$=1/8 in 
Y-123 or the Hg cuprates, in contrast to the case of
(La,Nd)-124; stripe domains in the latter system have longer-range order
($\geq 170\rm{\AA}$)\cite{Cuprates}.  

The authors acknowledge helpful comments from J. Ashkenazi, S. E. Barnes, 
M. Peter, and F. Zuo.  Work at the University of Miami was supported by NSF 
Grant No.~DMR-9631236, and at the University of Houston by NSF Grant 
No.~DMR-9500625 and the state of Texas through the Texas Center for 
Superconductivity.

\end{multicols}

\begin{references}

\bibitem{Cuprates} J. M. Tranquada {\it et al.}, Nature {\bf 375}, 561 (1995);
\prl {\bf 78}, 338 (1997).

\bibitem{DynamicStripes} J. M. Tranquada, ANL preprints, cond-mat/9702117
and cond-mat/9709325.

\bibitem{Hammel} P. C. Hammel {\it et al.}, \prb {\bf 57}, R712 (1998).

\bibitem{HamScal} P. C. Hammel and D. J. Scalapino, Phil. Mag. {\bf 74}, 523 (1996).

\bibitem{Loram} J. W. Loram {\it et al.}, \prl {\bf 73}, 1721 (1993);
J. W. Loram {\it et al.}, J. Supercond. {\bf 7}, 243 (1994).

\bibitem{PseudoGap} B. Battlogg {\it et al.}, Physica C {\bf 235-240}, 
130 (1994); B. N. Basov {\it et al.}, \prb {\bf 50}, 3511 (1994); 
J. L. Tallon {\it et al.}, \prl {\bf 75}, 4414 (1995); A. Loeser {\it et al.}, 
Science {\bf 273}, 325 (1996); G. V. M. Williams 
{\it et al.}, \prl {\bf 78}, 721 (1997).

\bibitem{UherRev} C. Uher, in {\it Physical Properties of High Temperature
Superconductors,} edited by D. M. Ginsberg (World Scientific, Singapore,
1993), Vol. III., p. 159.

\bibitem{Popoviciu} C. P. Popoviciu and J. L. Cohn, Phys. Rev. B {\bf 55}, 3155 
(1997); J. L. Cohn, {\it ibid.} {\bf 53}, R2963 (1996).

\bibitem{PfromTEP} J. L. Tallon {\it et al.}, Phys. Rev. B {\bf 51}, 12911 
(1995).

\bibitem{Krishana} K. Krishana {\it et al.}, \prl {\bf 75}, 3529 (1995).

\bibitem{CohnManganites} J. L. Cohn, C.-K. Lowe-Ma, and T. A. Vanderah, 
\prb {\bf 52}, R13134 (1995); J. L. Cohn {\it et al.}, \prb {\bf 56}, R8495 
(1997). 

\bibitem{muSRY123} Y. J. Uemura {\it et al.}, \prl {\bf 66}, 2665 (1991);
J. L. Tallon {\it et al.}, {\it ibid.} {\bf 74}, 1008 (1995).  

\bibitem{Occupancy} O. Chmaissem {\it et al.}, 
Physica C {\bf 217}, 265 (1993); E. V. Antipov {\it et al.}, {\it ibid.} 
{\bf 218}, 348 (1993); Q. Huang {\it et al.}, Phys. Rev. B {\bf 52}, 462 
(1995).

\bibitem{Samples} Q. M. Lin {\it et al.}, Physica C {\bf 254}, 207 (1996).

\bibitem{LSCOelastic} S. Sakita {\it et al.}, Physica B {\bf 219-220}, 216 
(1996); M. Nohara {\it et al.}, \prb {\bf 52}, 570 (1995).

\bibitem{VsoundMax} M. Peter and E. R. Walker, J. Appl. Phys. {\bf 48}, 
2820 (1977); B. L\"uthi, J. Mag. Mag. Mater. {\bf 52}, 70 (1985).

\bibitem{HgRaman} X. Zhou {\it et al.}, Phys. Rev. B {\bf 54}, 6137 (1996).

\bibitem{Jorgensen} J. D. Jorgensen {\it et al.}, \prb {\bf 41}, 1863 (1990).

\bibitem{NoteonMFP}  We take $C_L(100$K)=1.3$\times10^6$J/m$^3$K 
(Ref.~\protect{\onlinecite{Loram}}) and $v=3\times 10^3$m/s estimated
from dispersion curves [L. Pintschovius {\it et al.}, 
Physica C {\bf 185-189}, 156 (1991)].
\end{references}
\end{document}